# RPC with low-resistive phosphate glass electrodes as a candidate for the CBM TOF

A. Akindinov[a], V. Ammosov[b], V. Gapienko[b], Yu. Grishuk[a], F. Guber[c], N. Herrmann[d], O. Karavichev[c], S. Kiselev[a], A. Maevskaya[c], V. Razin[c], A. Semak[b], A. Smirnitskiy[a], Yu. Sviridov[b], V. Tiflov[c], K. Voloshin[a,*], V. Zaets[b], B. Zagreev[a]

[a]*Institute for Theoretical and Experimental Physics (ITEP), Moscow, Russia.*

[b]*Institute for High Energy Physics, Protvino, Moscow region, Russia*

[c]*Institute for Nuclear Research, Moscow, Russia*

[d]*Universität Heidelberg I. Physikalisches Institut der Universität, Philosophenweg 12, D - 69120 Heidelberg, Germany*

**Abstract**

Usage of electrodes made of glass with low bulk resistivity seems to be a promising way to adapt the Resistive Plate Chambers (RPC) to the high-rate environment of the upcoming CBM experiment. A pilot four-gap RPC sample with electrodes made of phosphate glass, which has bulk resistivity in the order of $10^{10}$ Ω cm, has been studied with MIP beam for TOF applications. The tests have yielded satisfactory results: the efficiency remains above 95% and the time resolution stays within 120 ps up to the particle rate of 18 kHz/cm$^2$. The increase in rate from 2.25 to 18 kHz/cm$^2$ leads to an increase of estimated 'tails' fraction in the time spectrum from 1.5% to 4%.

## 1. Introduction

The goal of the CBM[1] experiment, proposed at the future FAIR accelerator facility (GSI, Germany), is the study of strongly interacting matter under extremely high net baryon densities and moderate temperatures [1]. This as yet unexplored part of the QCD phase diagram is expected to be probed in fixed-target nucleus-nucleus collisions with beam energies between 10 and 40 A GeV. The CBM setup will have to deal with an event rate of 10 MHz, which, multiplied by a factor of about 1000 charged particles produced in each event, creates an unprecedented experimental challenge. In particular, all CBM sub-systems will be required to operate efficiently and consistently during the whole period of data taking in high radiation environment.

The Time-of-Flight (TOF) system of the CBM experiment is proposed to be assembled of Resistive Plate Chambers (RPC), which by now have become a reliable and proven tool for timing measurements [2–5]. The TOF wall will be positioned about 10 m apart from the target, cover about 100 m$^2$ and provide the timing accuracy of less than 100 ps with almost 100% detection efficiency. There are a number of technical parameters that remain to be addressed, such as the high-rate capability, long-term stability and reliable overall performance of the TOF system.

---

* Corresponding author.
Email address: `Kirill.Voloshin@itep.ru` (C. Author).

[1] Compressed Baryonic Matter.

## 2. CBM TOF requirements for RPC

RPC, normally used for TOF applications, are multi-gap chambers operated in the saturated avalanche mode. They typically consist of several planar glass plates, 0.5–2 mm thick, which are kept apart at a fixed distance of 0.2–0.4 mm. A high and uniform electric field of 8–11 kV/mm applied between the electrodes yields immediate avalanche amplification of initial ionization produced in the gas. Time resolution and detection efficiency of multi-gap RPC (MRPC) as well as dependence of these characteristics on main chamber parameters have been subjected to careful studies with MIP beams [6,7]. Tested detectors have demonstrated excellent TOF performance, which combines a wide efficiency plateau (with efficiency staying close to 100%), high timing resolution (confidently below 100 ps) and a wide streamer-free range in the high voltage. Improvement of the MRPC time resolution with increasing number of gaps and electric field strength has been observed.

To adapt the CBM TOF system design for the tasks of the experiment, simulations of physical events have been performed in the framework of the CbmRoot [8] software, which is based on the GEANT3 [9] and Root [10] packages. For simplicity, the TOF wall was described as a $10 \times 10$ m$^2$ plane with a $0.8 \times 0.8$ m$^2$ hole in the middle, positioned 10 m apart from the target. The RPC filling was represented as layers of Al–Gas–Glass–Gas–Al with corresponding thicknesses of 0.2–0.12–0.54–0.12–0.2 cm. Default descriptions of other CBM sub-systems had been already included in CbmRoot.

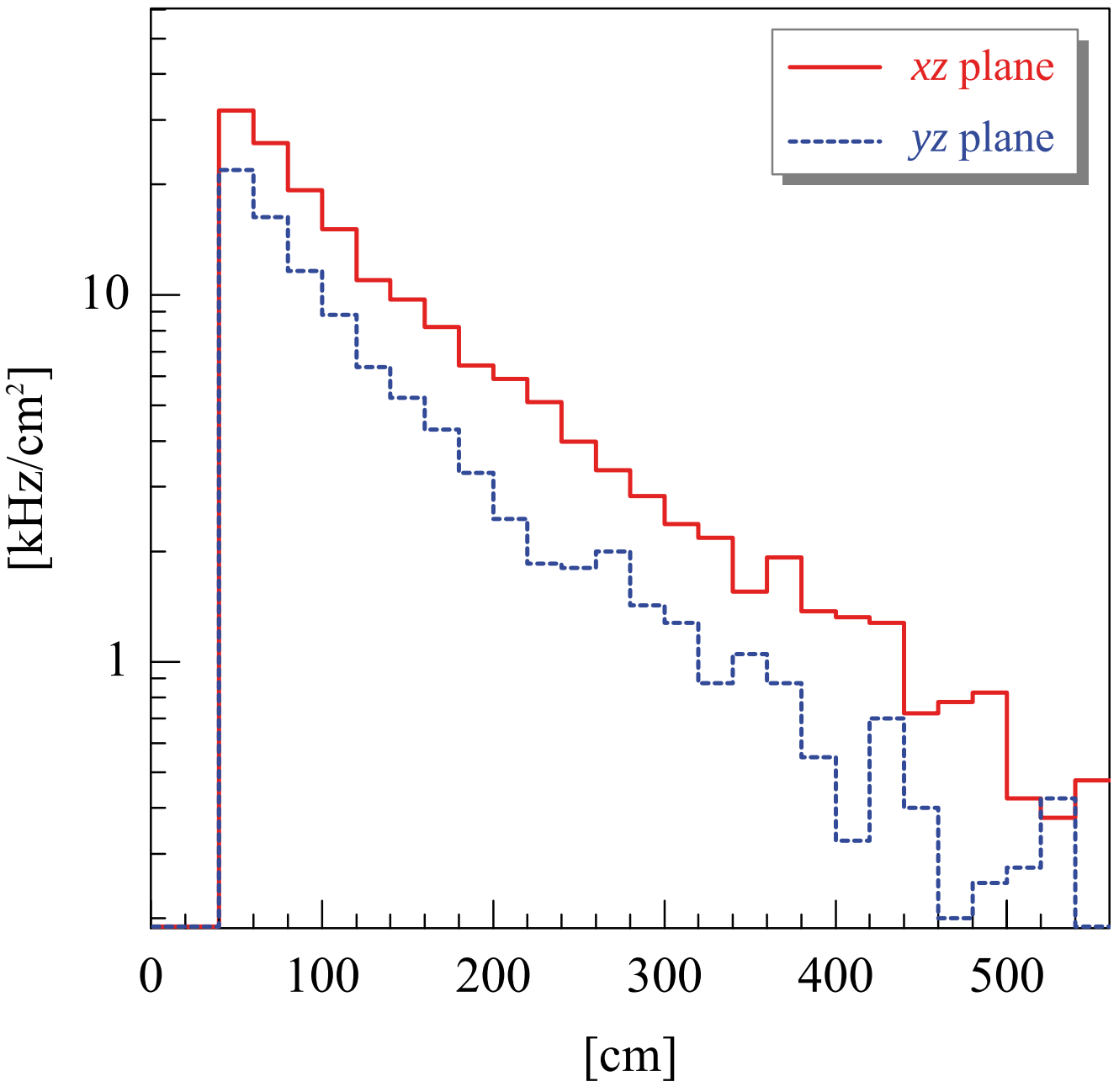

Fig. 1. Simulated distributions of hit rate density in the TOF plane in Au+Au events at 25 A GeV (the *x* axis is perpendicular to the magnetic field direction).

Assuming the expected event rate at about 10 MHz, minimum bias Au+Au events at 25 A GeV were generated within the UrQMD model [11]. Simulation results comprise both primary particles coming from the interaction point as well as secondaries produced in the material of the detector sub-systems. The resulting distribution of hit rate density in the TOF plane is presented in Fig. 1, where the projections are made onto orthogonal coordinates, so that the *xz* plane is perpendicular to the magnetic field. The overall hit rate density for small emission angles of 50–100 mrad (corresponding to the central area of the TOF wall within the radius of 50–100 cm) was estimated at the level of 20 kHz/cm$^2$. This value determines the limitation of the rate capability for the CBM TOF system.

The TOF detector occupancy is important to the choice of the RPC cell size. Shown in Fig. 2 is the simulated density of charged tracks in the TOF plane in a single central collision. In the vicinity of the beam pipe the track density increases up to about 1 track/dm$^2$. Under such conditions, an acceptable occupancy of 10% can be established if the size of the basic RPC cell is about 10 cm$^2$. This value determines the granularity limitation for the CBM TOF system, at least in its innermost part.

## 3. Low-resistive phosphate glass for RPC electrodes

The RPC rate capability is limited by the time interval needed for a localized discharge to dissolve from the glass electrode. With all the other parameters of RPC being fixed, this time is determined by the bulk resistivity of glass. The resistivity of conventional glass ($\sim 10^{13}$ Ω cm) results in the rate capability of RPC being as low as several hundred Hz/cm$^2$, which is unacceptable for the CBM TOF system. Usage of low-resistive glass with the resistivity of $10^{10}$–$10^{11}$ Ω cm is an attractive way of improving the RPC rate capability.

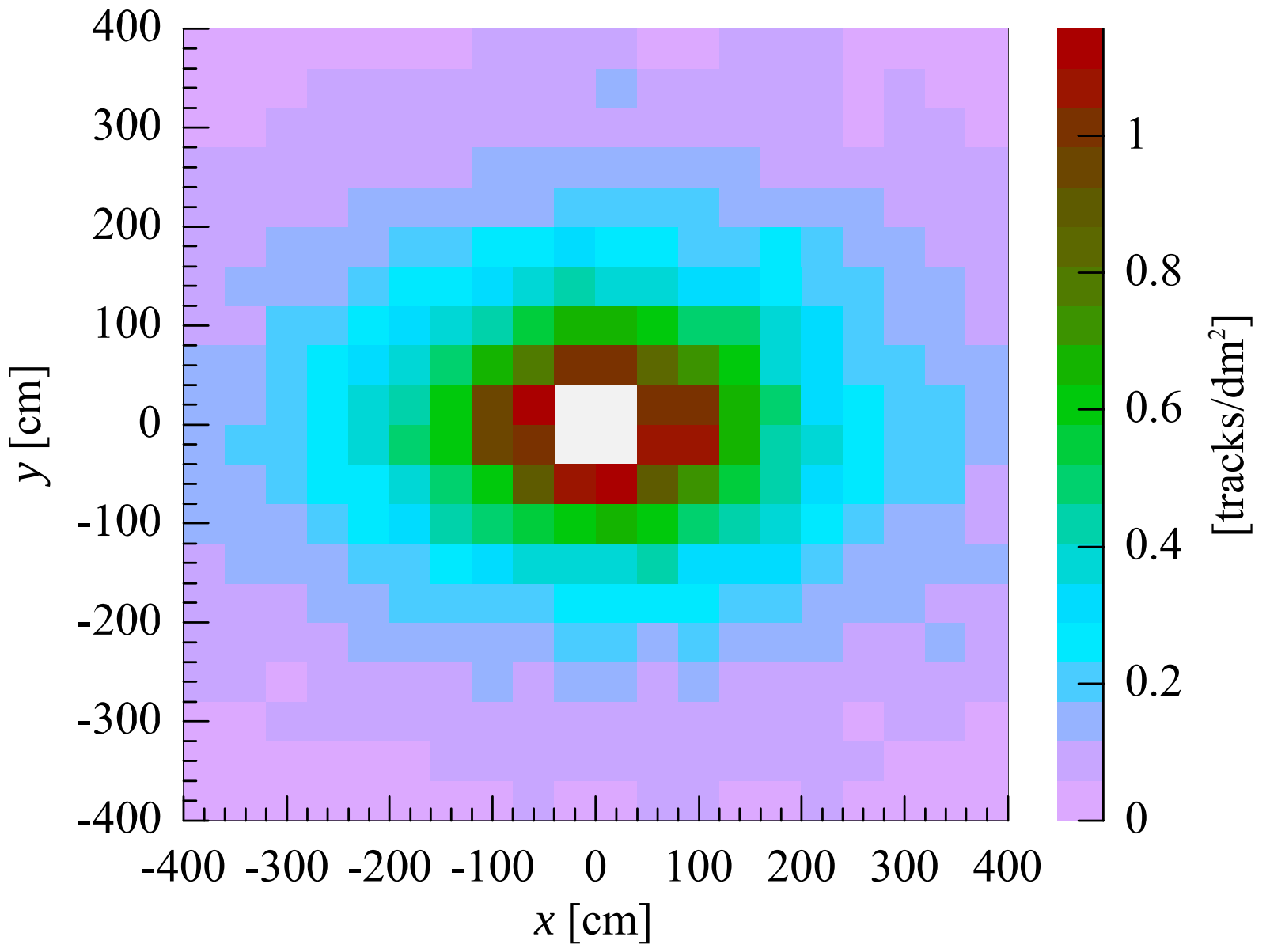

Fig. 2. Simulated density of charged tracks in the TOF plane in a single central Au+Au event at 25 A GeV.

There is presently no commercially available window glass with the bulk resistivity of less than $10^{12}$ Ω cm. One of the most promising materials for the RPC electrodes is a semiconductive glass, of which there are three known types: phosphate, silicate and borosilicate. All three contain oxides of transitional elements, possess electron type of conductivity and have a relatively low bulk resistivity of $10^8$–$10^{11}$ Ω cm. They are black in color and opaque for the visible light. As compared to window glass, technology of the semiconductive glass production is more complicated, the resulting glass resistivity is sensitive to the chemical composition of raw material and glass melting procedure.

The phosphate glass, used in the pilot RPC (described later in Section 4), was molded into samples sized 100 × 100 mm$^2$, several mm thick. These samples were then filed down to the thickness of 2 mm and polished. The measured dependence of glass bulk resistivity on the applied voltage is presented in Fig. 3. With voltage being within the limit of several hundred volts, the resistivity was found to be in the order of $10^{10}$ Ω cm.

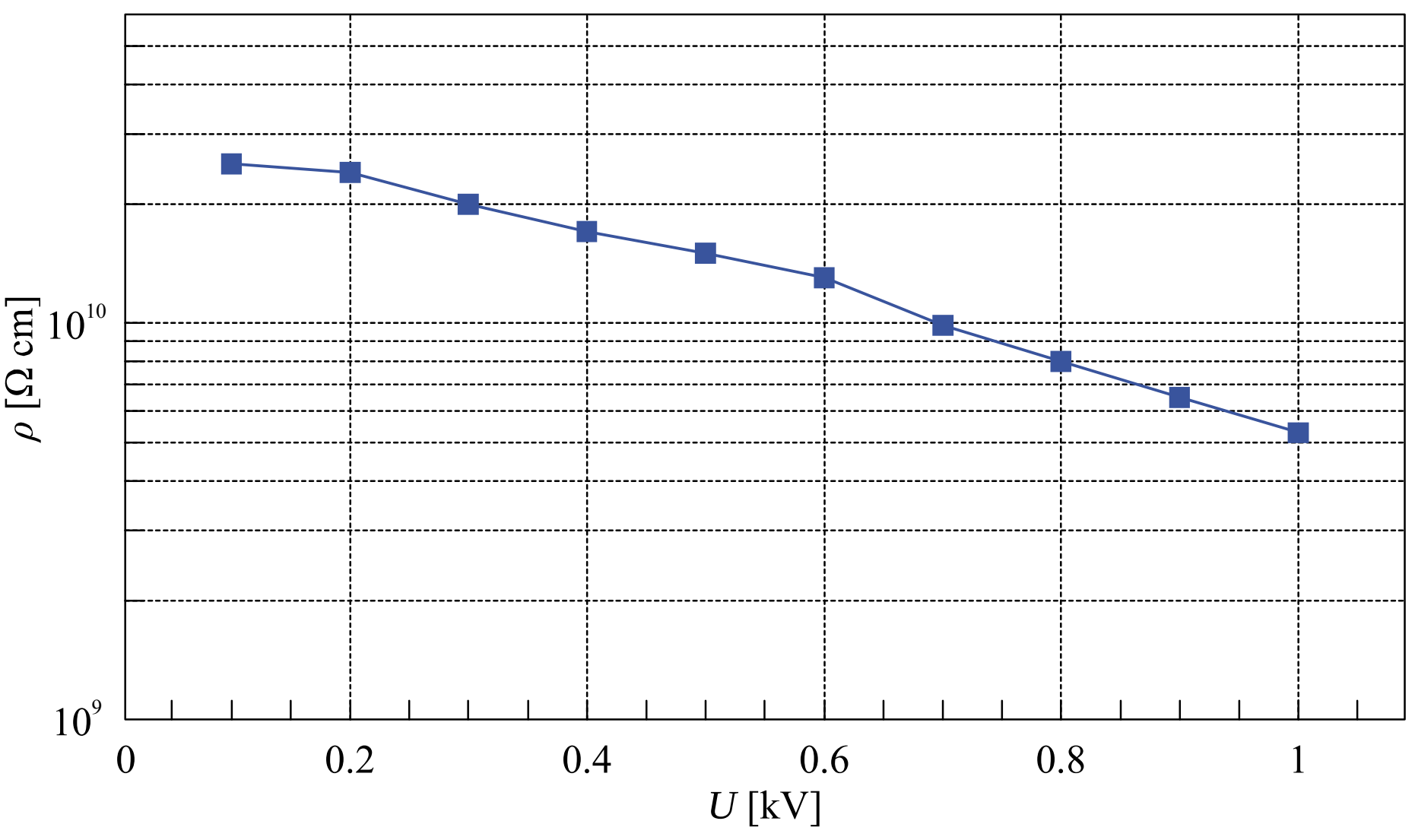

Fig. 3. Dependence of the bulk resistivity of phosphate glass on high voltage.

## 4. RPC prototype description

A sketch of the tested RPC prototype with electrodes made of phosphate glass is shown in Fig. 4. Six cut and polished glass plates, sized 50 × 55 × 2 $mm^3$, were arranged in two double gap stacks with the intermediate glass plates left electrically floating. Gap sizes (0.3 mm) were fixed with fishing-line spacers, which ran directly through the RPC working area. Carbon resistive tapes were used to apply high voltage to the stack. Signal pulses could be taken either from both anode and cathode pads with the sensitive areas of 40 × 40 $mm^2$ (differential readout) or exclusively from the central anode pad (single readout). The dark rate was found to be about 2100 Hz.

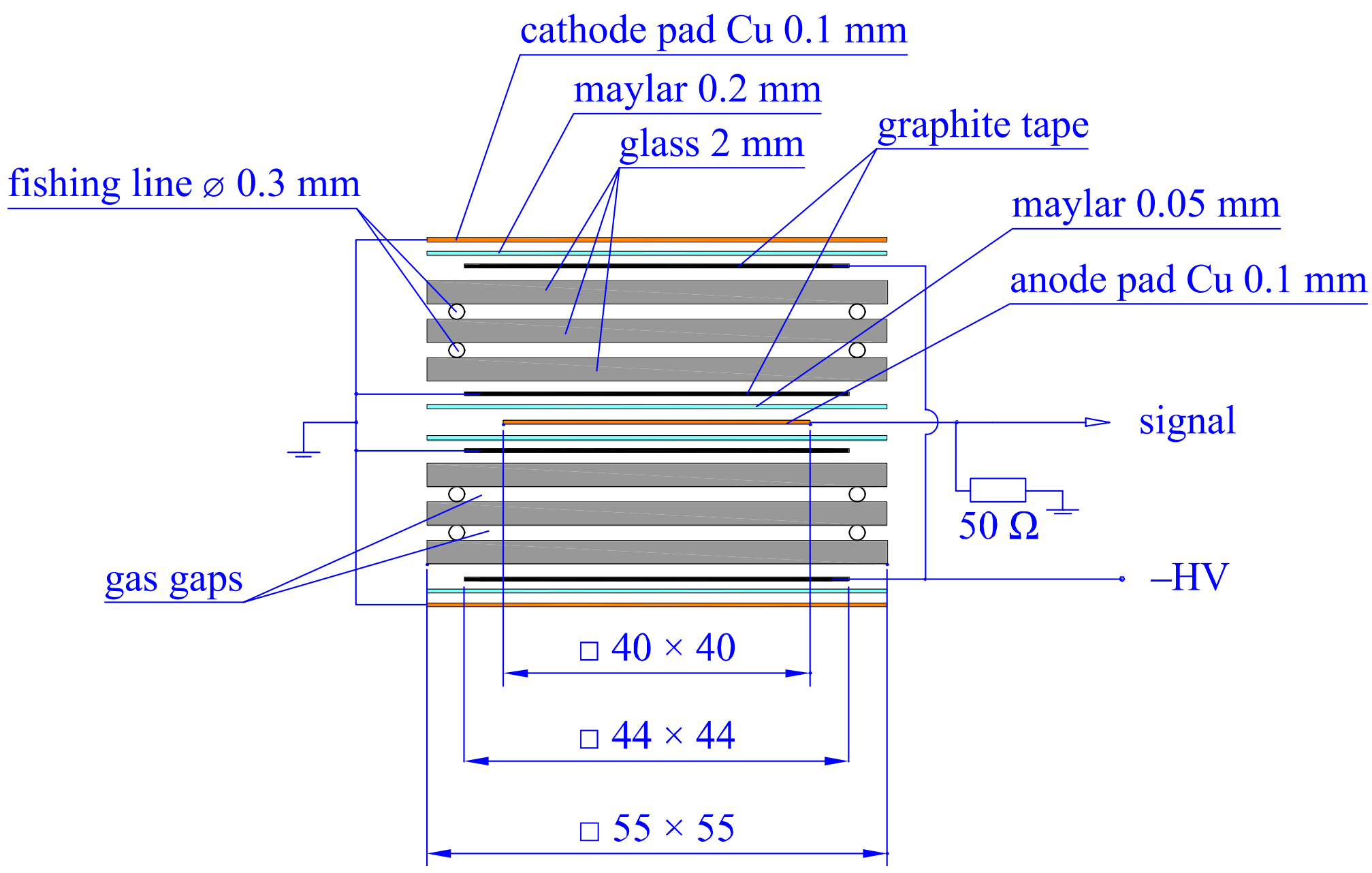

Fig. 4. Schematic drawing of RPC with electrodes made of phosphate glass.

## 5. RPC tests with beam

Beam tests of the chamber described above were performed with 1.28 GeV/*c* $\pi^-$-mesons at the ITEP PS. The RPC gas filling was mixed of 85% $C_2H_2F_4$ + 5% iso-$C_4H_{10}$ + 10% $SF_6$. The beam focusing could be adjusted with a collimator, the corresponding range of RPC rate density being 0–18 kHz/$cm^2$. Monitored with a MWPC, the beam profile was found to stay uniform over the area of 5 × 5 $mm^2$ in the RPC plane. No means to monitor the temporal bunch uniformity were available.

The size of the trigger area was 1.8 × 1.8 $cm^2$, determined by two scintillation counters positioned cross-wise before the tested RPC. The same counters produced the start signal for timing measurements, the time resolution of each counter being about 50 ps. The MIP events were triggered with these two counters in coincidence with a beam monitor consisting of two larger scintillation counters placed before and after the RPC. The beam monitor was also used to calculate the RPC rate.

The front-end electronic channel was based on a commercially available MAX 3760 amplifier[2] with the following parameters:

- gain — 2.5 V/pC

[2] Produced by Maxim Integrated Products, Inc., http://www.maxim-ic.com

- rise-time (with RPC connected) — 1.4 ns
- noise — 1200 $e^-$ + 300 $e^-$/pF
- linearity — up to 100 fC

The amplifier noise contribution into the overall detector time resolution was found to be less than 20 ps.

After amplification, the RPC signals were split for independent charge and timing measurements. Signal charges $Q$ were measured with a charge-sensitive Le Croy QDC 2249 with a 50 ns gate. Time $T$ was measured with a discriminator coupled to a Le Croy TDC 2228. Data readout and further processing were performed by means of an IBM PC. To make the slewing $T$–$Q$-correction, a polynomial fit was applied to the measured event distribution in the time-charge coordinates. The range of the slewing correction along the $T$-axis was as large as 1 ns. The corrected data was then used to calculate the RPC timing parameters.

## 6. Results and discussion

The results of efficiency and time resolution measurements under different rates in the range of 2.25–18 kHz/cm$^2$ are summarized in Fig. 5. The increase in rate leads to a degradation of efficiency from 98% to 95%, while the time resolution gradually deteriorates from 90 ps to 120 ps.

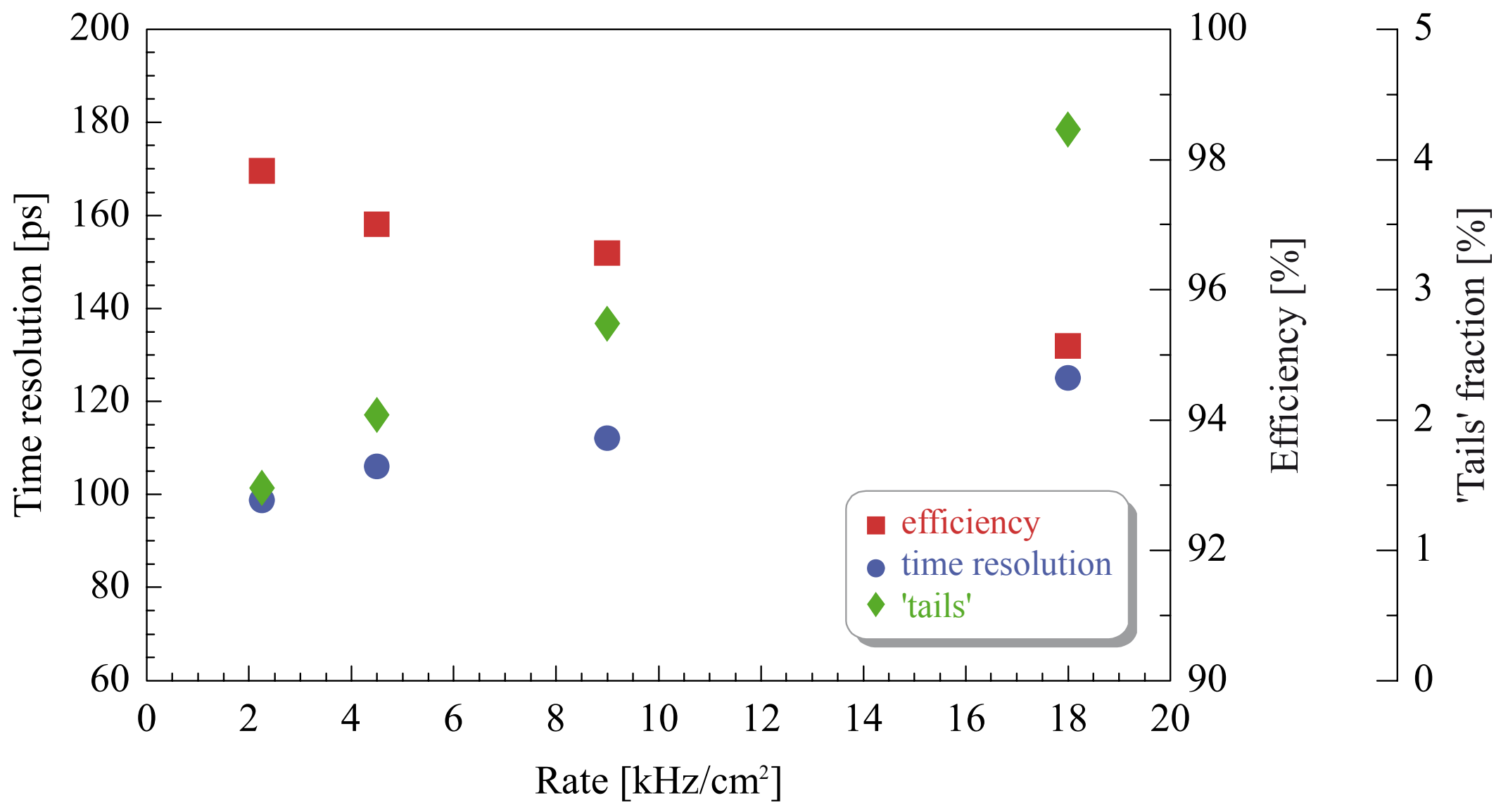

Fig. 5. Time resolution (squares), efficiency (circles) and estimated ‘tails’ fraction in the timing spectrum (diamonds) of RPC as functions of the rate.

To illustrate this effect, Fig. 6 shows charge (Fig. 6a) and $T$–$Q$-corrected timing spectra (Fig. 6b) of the tested RPC corresponding to four rate values. The essential statistical information and fit results may be found directly on the plots. The RPC time resolution is traditionally calculated by fitting the corrected timing spectra with Gaussian distribution and subsequent quadratical subtraction of the time resolution of the start system from the Gaussian sigma. However, the true shapes of the timing spectra are not exactly Gaussian, they have ‘tails’ on both sides, and this distinction becomes more visible with increasing rate density. The ‘tails’ impair the actual time resolution and as a result weaken the RPC performance in terms of precise particle identification.

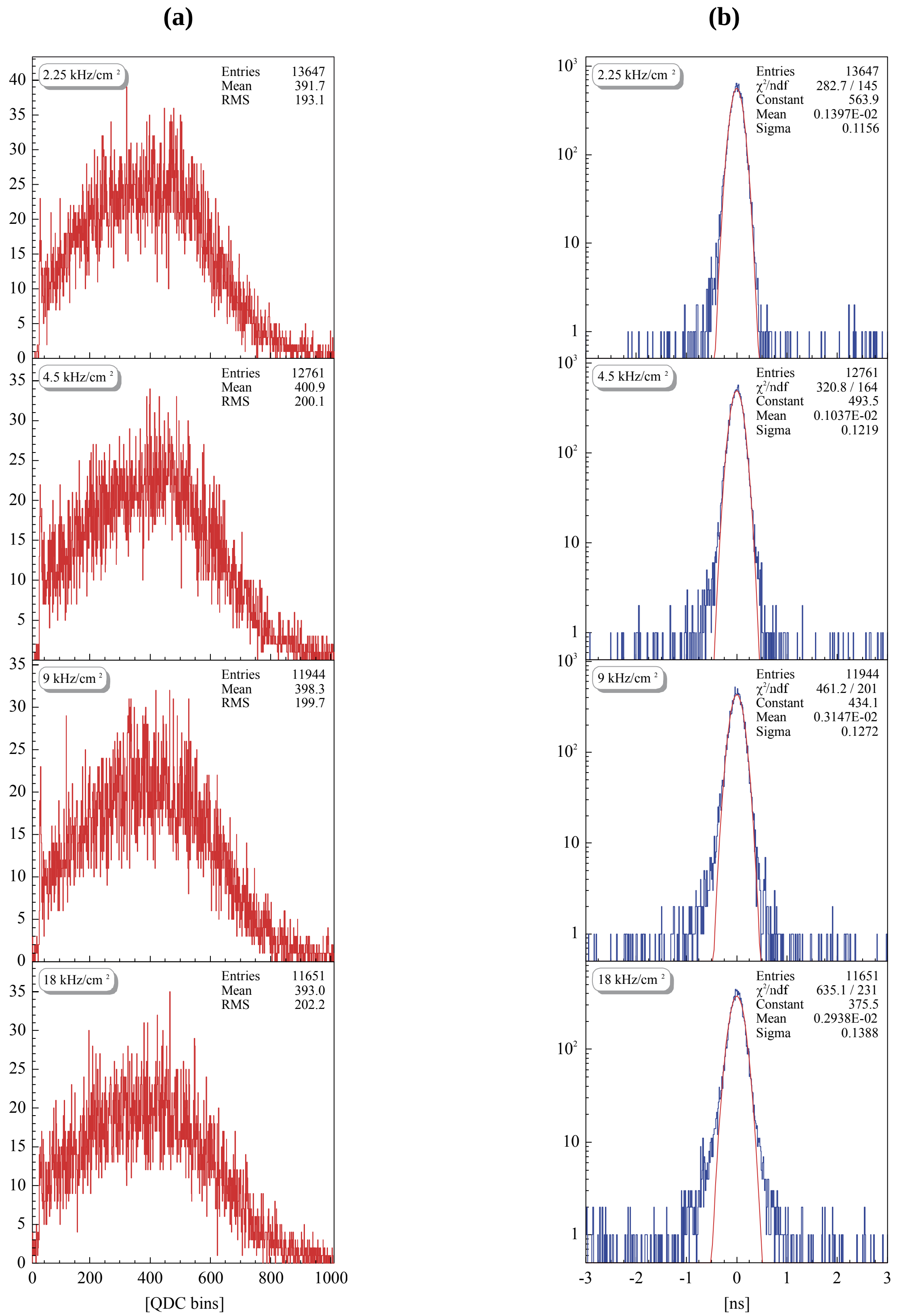

Fig. 6. Measured RPC characteristics under different rates: (a) charge and (b) *T*–*Q*-corrected timing spectra. The rate values and statistical information are printed on the plots.

The ‘tails’ fraction may be estimated as $tail = 1 – n_{3\sigma}/n$, where $n$ and $n_{3\sigma}$ are respectively the total number of effective events in the timing spectrum and the number of events within the three-sigma limit of the Gaussian fit (such an estimate is rough since it significantly depends on the *T–Q*-correction and fitting results). The calculated ‘tails’ fraction is plotted in Fig. 5, where one can see that this fraction tends to grow linearly with rate density. On the other hand, the change of rate does not affect the charge spectra of RPC, as shown in Fig. 6a. It is hence reasonable to assume that the ‘tails’ are caused by a pile-up of signals from successive events, which is a pure electronic effect. Further R&D is required to confirm this assumption and find a way to diminish the ‘tails’ to the least possible level.

With the discriminator threshold set at 0.2 mV, the knee of the HV plateau for the RPC efficiency (defined as the point in HV where efficiency exceeds 95%) has been found at 5.6 kV. The noise level at this point is about 1 Hz/cm$^2$. The noise then rapidly increases with increasing HV, surpassing 100 Hz/cm$^2$ at HV ≈ 6.2 kV and leading to the unstable behavior of the chamber. This effect may signify that phosphate glass is probably not the best solution for the high-rate RPC, and other types of low-resistive glass have to be investigated in the near future.

## 7. Conclusion

Simulations predict that the innermost part of the CBM TOF system, proposed to be assembled of RPC cells, will be operated under extreme hit rates of about 20 kHz/cm$^2$. To improve the high-rate capability of RPC, usage of electrodes made of low-resistive glass was suggested. Phosphate glass with the bulk resistivity in the order of $10^{10}$ Ω cm was the first such glass to be tested.

A multigap RPC prototype of the minimal (4 × 0.3 mm gaps) configuration with electrodes made of phosphate glass has been tested with a beam and has shown satisfactory behavior in terms of efficiency and time resolution. The increase in rate from 2.25 to 18 kHz/cm$^2$ leads to acceptable degradations in efficiency (from 98% to 95%) and in time resolution (from 90 ps to 120 ps). However, the fraction of ‘tails’ in the RPC timing spectrum grows with increasing rate density from 1.5% to 4%. This feature currently represents the main disadvantage of such a chamber. Moreover, the noise rate of the tested RPC has shown an undesirably strong dependence on the high voltage applied.

Despite the observed degradation of RPC performance under high rate densities, the results presented are a significant step forward in the development of detectors for high-rate operations and call for further R&D.

## Acknowledgements

We kindly thank A. Golovin and N. Mishina for their efforts to produce the RPC prototype. We are grateful to the members of the CBM Collaboration, especially to P. Senger and W.F.J. Mueller, for fruitful discussions and their valuable remarks.

This work was partially supported by the INTAS grant #03–54–3891.